\documentclass[pre,superscriptaddress,twocolumn,a4paper,showpacs]{revtex4-1}
\usepackage{graphics}
\usepackage{amssymb}
\usepackage{wasysym}
\usepackage{latexsym}
\usepackage{graphicx}
\usepackage{epsfig}
\usepackage{subfigure}
\begin{document}

\title{Wisdom of crowds: much ado about nothing}  

\author{Sandro M. Reia}
\affiliation{Instituto de F\'{\i}sica de S\~ao Carlos,
  Universidade de S\~ao Paulo,
  Caixa Postal 369, 13560-970 S\~ao Carlos, S\~ao Paulo, Brazil}

\author{Jos\'e F.  Fontanari}
\affiliation{Instituto de F\'{\i}sica de S\~ao Carlos,
  Universidade de S\~ao Paulo,
  Caixa Postal 369, 13560-970 S\~ao Carlos, S\~ao Paulo, Brazil}

\begin{abstract}
The puzzling idea that the combination of independent estimates of the magnitude of a quantity  results in  a very accurate prediction, which is  superior to any or, at least, to most of the  individual estimates is known as the  wisdom of crowds. Here we use  the Federal Reserve Bank of Philadelphia's Survey of Professional Forecasters   database  to  confront the statistical and psychophysical  explanations of this phenomenon.  Overall we find that the data do not support  any of the proposed explanations of the wisdom of crowds. In particular, we find a positive correlation between the variance (or diversity) of the estimates and the crowd error in disagreement with some  interpretations of the diversity prediction theorem. In addition,  contra the predictions of the psychophysical  augmented quincunx model, we find that the skew of the estimates offers no information about the crowd error. More importantly,  we find that the crowd beats all individuals in less than 2\% of the forecasts  and beats most individuals  in less than 70\% of the forecasts, which  means that there is a sporting chance that an individual selected at random will perform better than the crowd. These results contrast starkly with  the performance of  non-natural crowds composed of  unbiased forecasters which beat most individuals in practically all forecasts.    The moderate statistical advantage of a real-world crowd over its members  does not justify the ado about its wisdom, which is most likely a product of the selective attention fallacy.
%
\end{abstract}

\maketitle

\section{Introduction}\label{sec:intro}

The  wisdom of crowds usually refers to the notion that  a  collection of individuals  -- the crowd -- can solve problems better than most individuals within it, including experts \cite{Surowiecki_04}.   The idea  was  brought to light by Galton's 1907  analysis of a contest to guess the weight of an ox  at the West of England Fat Stock and Poultry Exhibition in Plymouth \cite{Galton_07,Wallis_14}.  Despite  being more than a century old,
the  wisdom of crowds is still a  subject of  fascination  for   laypeople   and  scientists as well.   
This fascination  stems from  reports of the remarkably accurate appraisal  produced by  the statistical average of  independent estimates of the magnitude of an unknown quantity.  For instance, in the ox-weighing contest,  the crowd overestimated the weight of the ox by less than 1\% of the true weight \cite{Galton_07}.

There are a few attempts to explain the wisdom of crowds using purely statistical arguments. The first idea that comes to mind is that  the individual estimates are unbiased, that is, that the errors spread in equal proportion around the true value of the unknown quantity so that they cancel out when the estimates are  combined together \cite{Bates_69,Armstrong_01}. Although it is hard to accept the nonexistence of systematic errors on the individual estimates, this explanation is rather popular  perhaps because its underlying   assumption is difficult to verify \cite{Sunstein_06}. 

A  somewhat more sophisticated explanation for the wisdom of crowds  
 is offered by the  diversity prediction theorem  \cite{Page_07}, which  asserts that the  error of the collective estimate is never greater  than the average individual  error.   Moreover, as  hinted by its name,  the  theorem has a say in the role of the diversity of the individual estimates.  In fact, since the theorem asserts  that the  quadratic collective error  equals the average quadratic individual error minus the diversity of the estimates, one is tempted to think that the increase of the diversity would improve the crowd accuracy  \cite{Page_07}.

In addition to the statistical explanations, there is a psychophysical model  of the wisdom of crowds, viz., the augmented quincunx  model of probabilistic cue categorization \cite{Nash_14},  which assumes that there exists a typical value of the estimated quantity   that  is common knowledge of the population. The real object, say Galton's  ox,  is then compared with the prototypical object through a number of perceptible cues that may be categorized  incorrectly by the individuals. The  main prediction of the model is that information about the collective error can be inferred from the  skewness of  the  distribution of  the estimates \cite{Nash_14}.

In order to test the predictions, as well as the relevance, of the aforementioned  explanations for the wisdom of crowds,   we  use forecasts of economic indicators that  are publicly available   in the Federal Reserve Bank of Philadelphia's (FRBP)  Survey of Professional Forecasters  \cite{FRBP}. As this database offers quarterly projections of the economic indicators, we can tune the difficulty of the forecasts by controlling for the forecast range. We find that the collective error and the diversity  of the estimates are  affected significantly by that range.  Most interestingly,  we find that the difficulty of the forecast is typically associated with a large variance and a long left tail of the distribution of estimates.

  Overall we find that the data do not support  any of the proposed explanations for the wisdom of crowds. First, in opposition to the unfounded interpretation of  the diversity prediction theorem, we  find a positive correlation between the collective error and the diversity of estimates. Second, we find that, once the range of the forecasts is accounted for, the skewness of the distribution of estimates    does not  influence significantly  the  crowd accuracy, in opposition to  the predictions of the augmented quincunx model.  Third, we find that the  unbiased estimates assumption confers on the crowd  an enormous advantage over the individuals within it,  which is at great variance with the data.


Our main finding, which was obtained using almost $10^4$ forecast experiments for several economic indicators in  the  FRBP forecast database, is that only  rarely the crowd beats all individuals within it. More precisely, this happens in less than 2\% of the forecast experiments that we analyzed.  In addition, the crowd beats most individuals within it in less than 70\% of our experiments, which means that there is a fair chance that an individual selected at random will perform better than the crowd. Clearly, the purely statistical advantage of the crowd over its members  does not justify the high esteem it enjoys, which is most likely a product of the selective attention fallacy.

The rest of this paper is organized as follows. In section \ref{sec:model} we offer  an outline of the main ideas used   to explain the wisdom  of crowds, viz., the diversity prediction theorem \cite{Page_07}, the augmented quincunx model \cite{Nash_14} and the unbiased estimates assumption \cite{Bates_69}.   In  section \ref{sec:data} we describe briefly the FRBP  forecast database \cite{FRBP} from where we have extracted the forecast experiments.
In  section \ref{sec:res} we present and analyze the results of those experiments in the light of the  known explanations for the wisdom of crowds. In   \ref{AppA} we replicate those forecast experiments using virtual unbiased agents so as to verify the predictions  of the unbiased estimates assumption. In   \ref{AppB} we revisit a few wisdom-of-crowds experiments where the participants are laypeople and show that their results are consistent with the results obtained  using the expert forecasters of the FRBP   database.
Finally, section \ref{sec:disc} is reserved to our concluding remarks.

\section{Three frameworks for understanding the wisdom of crowds}\label{sec:model}

Here we describe briefly three frameworks that claim to explain the  wisdom of crowds, viz., the diversity prediction theorem, the augmented quincunx model and the unbiased estimates assumption. These proposals make a variety of specific predictions that will be tested using the FRBP forecast database in section \ref{sec:data}.

\subsection{The diversity prediction theorem}\label{sec:DPT}

The diversity prediction theorem  is viewed as a  main achievement to  those allured by  the idea that  the performance of groups can be boosted by increasing the diversity of their members  \cite{Page_07}, although even simple agent-based models indicate that the effects of diversity can be rather unpredictable in nontrivial  problem-solving  scenarios \cite{Fontanari_16}. This theorem  shows that the quadratic collective error  is related in a very simple manner to the average  quadratic individual error  and to a measure of the diversity of the estimates. More pointedly, let us denote by $g_i$  the estimate of some unknown quantity  by individual  $i=1, \ldots, N$,  and by $G$  the true value of the unknown quantity. The collective estimate is defined as the arithmetic mean of the individual estimates, that is,  
\begin{equation}\label{av}
 \langle g \rangle =  \frac{1}{N} \sum_{i=1}^N g_i, 
\end{equation}
as usual. We note, however,  that Galton used the median of the individual estimates as the crowd estimate  in his seminal  ox-weighing experiment \cite{Galton_07},  though the arithmetic mean proved to be a much better estimator in that case \cite{Wallis_14}.
Thus,   the collective error is defined as  the signed quantity
\begin{equation}\label{gamma}
\gamma =   G - \langle g \rangle .  
\end{equation}
  In most of this paper, however, we consider  the unsigned collective error  $| \gamma |$; the only exception is the discussion of  the  prediction of  the augmented quincunx model that  the signs of $\gamma$ and of the skewness of the distribution of estimates are negatively correlated.
Defining the average quadratic individual error as
\begin{equation}\label{eps}
\epsilon = \frac{1}{N} \sum_{i=1}^N \left ( g_i - G \right )^2
\end{equation}
and the diversity or  variance of the estimates as
\begin{equation}\label{delta}
\delta = \frac{1}{N} \sum_{i=1}^N \left ( g_i  - \langle g \rangle  \right )^2,
\end{equation}
we have the identity
\begin{equation}\label{DPT}
\gamma^2 =  \epsilon - \delta ,
\end{equation}
which  is Page's diversity prediction theorem \cite{Page_07}. It  asserts  that the  quadratic collective error  equals the average quadratic individual error minus the prediction diversity.  This result is sometimes viewed as indication that the increase of the prediction diversity $\delta$ results in the decrease of the quadratic collective error $\gamma^2$. Of course, since $\delta$ and $\epsilon$ cannot be varied independently of each other,  this interpretation is not correct.  Nevertheless, the relationship, if any, between the  diversity of the estimates  and the collective error is a very interesting issue that can be investigated using a large number of equivalent  forecast experiments, which is the approach we follow in this paper. We note that the diversity of estimates $\delta$ is known in the statistical literature as the precision of the estimates, that is, the closeness of repeated estimates (of the same quantity)  to one another \cite{Tan_14}.

We stress that the diversity prediction theorem guarantees only that the quadratic collective error $\gamma^2$ is never greater than the mean quadratic  individual error $\epsilon$. As we will see in section \ref{sec:data}, this is not a very useful result from the practical perspective since, for instance, it does not imply that  the crowd  is better than most individuals. In particular, we will report an experiment where the collective estimate is worse than the estimate of about 85\% of the individuals. Hence,  $\gamma^2 \leq \epsilon$ does not imply that it is always advantageous to favor the collective estimate over  the estimate of a randomly chosen individual in the group. 

As an amusing side note,  we mention the resemblance between the discussions about the value of the diversity prediction theorem and the arguments about the relevance of the celebrated Price equation for evolutionary biology  \cite{Price_70}. We note that  Price's equation, which has a straightforward derivation from the definition of fitness,  is considered by many researchers  as a mere mathematical tautology  that has no predictive value at all \cite{Frank_12}.

It is not possible to investigate the influence of the diversity  of the estimates on the collective  error  using a single forecast experiment  since the values of $\gamma$, $\epsilon$ and  $\delta$ are fixed for a particular experiment. The solution is to consider a large ensemble of roughly equivalent experiments and to look at the correlations between those quantities. This can be achieved artificially by selecting random subsamples of the estimates of a single experiment to produce many virtual experiments with fewer estimates than the original one \cite{Davi_20}. The problem with that approach is that the resulting virtual experiments  are not independent.  Here we use the FRBP forecast database to collect the  independent forecast experiments necessary for the correlation analysis.  We note that since  the collective estimate  $\langle g \rangle$ and the true value $G$ may  have different values for different experiments, it is necessary to introduce the  dimensionless quantities $ \gamma/G$, $\epsilon^{1/2}/G$ and $\delta^{1/2}/\langle g \rangle$ to properly compare the experiments.

\subsection{The augmented quincunx model of judgment}\label{sec:AQ}

The augmented quincunx  is a psychophysical model of probabilistic cue categorization \cite{Nash_14},  whose name was inspired by a probability device  invented by Galton in 1873 to demonstrate the central limit theorem \cite{Stigler_89}. The basic idea behind this model is the assumption that there is a typical value of the unknown quantity that is common knowledge gained through experience. For instance, in the ox-weighing experiment, it is assumed that the population (or at least the participants of the contest) share the knowledge that  the typical weight of oxen is $\hat{G}$.   In order to estimate the weight $G$  of a  particular ox,  the contestants focus on a number of perceptible cues
$c=1, \ldots, C$  that are correlated with $G$ (e.g., the height of the ox or the degree to which its ribs are showing). If  a cue $c$ indicates that the ox is heavier (lighter) than the prototype ox then the typical weight is increased (decreased) by a factor $\eta_c$. Hence 
\begin{equation}\label{AQ1}
G = \hat{G} + \sum_{c=1}^C \eta_c ,
\end{equation}
where $\eta_c $ can be positive or negative depending on the correlation between cue $c$ and the ox weight. 
Stochasticity enters the augmented quincunx model because the contestants can perceive a cue incorrectly. More pointedly,  the estimate of contestant $i$ is 
\begin{equation}\label{AQ2}
g_i = \hat{G} + \sum_{c=1}^C u_c \eta_c ,
\end{equation}
where $u_c$ is a random variable that takes on the value $+1$ with probability $p$ and the value $-1$ with probability $1-p$  \cite{Nash_14}. This means that a cue is perceived correctly  with probability $p$ and  incorrectly with probability $1-p$.
 In particular,  if   individual $i$ can perceive all cues correctly (i.e., $p=1$) then its estimate  is perfect (i.e., $g_i = G$), despite the fact that  $\hat{G} \neq G$.  For simplicity, the model  assumes that all contestants are equivalent, i.e., the cue categorization probability $p$ is the same for all individuals.

Although  $\hat{G}$, $C$  and $\eta_c$ are hard to access quantities, the augmented quincunx model makes some  interesting general predictions. For instance, the model predicts that  the collective error $\gamma$ can be inferred from the degree and direction of the skewness of the distribution of individual  estimates: estimate distributions have greater negative (positive) skew when the mean estimate $\langle g \rangle$  underestimates (overestimates) the true value $G$ by greater margin  \cite{Nash_14}. 
 This claim means that there is a negative  correlation between  $\gamma/G$ and  the skewness $ \mu_3 $. We recall that the skewness $\mu_3$ of a distribution is  a  dimensionless measure of its asymmetry, which is defined as
\begin{equation}\label{skew}
\mu_3 =  \frac{1}{N} \sum_{i=1}^N \left (\frac{ g_i - \langle g \rangle}{\delta^{1/2}} \right )^3,
\end{equation}
where $\delta$ is the sample variance  of the estimates defined in equation (\ref{delta}) and $\langle g \rangle$ is the mean of the estimates defined in equation (\ref{av}).  A negative value of $\mu_3$ implies that the left tail of the distribution of estimates is longer than the right tail, whereas a positive $\mu_3$  indicates a  right-tailed distribution.

\subsection{The unbiased estimates assumption}\label{sec:UF} 

A  natural explanation for the wisdom of crowds  involves  the  well-known fact that the  combination of unbiased and independent
estimates   guarantees  the accuracy of the statistical average, provided  the number of estimates is large (see, e.g., \cite{Bates_69,Armstrong_01}). In other words, if the estimates made by numerous different people scatter symmetrically  around the truth, then the collective estimate is likely to be very accurate.  Of course, the trouble with this explanation is the assumption that  the individual estimates are unbiased, that is, that their means coincide with the true value of the unknown quantity. If correct,  this assumption would imply  that  one could harvest the benefits of the  wisdom of crowds by asking a single individual to make several  estimates at different times (see, e.g.,  \cite{Vul_08}).  The unbiased estimates assumption is the limit of 
the augmented quincunx model where $ \hat{G} = G$ and $p=1/2$.

 In \ref{AppA} we  use the unbiased estimates assumption as a null hypothesis  and replicate our analysis of the FRBP forecast experiments by replacing the expert economists by   virtual unbiased  forecasters. We  stress, however, that    ordinary people are not  unbiased forecasters.  On the contrary,  unbiased forecasters are the ultimate experts because their estimates fluctuate symmetrically (on the average) around the true value of the quantity being estimated, i.e., they somehow `know'  its true value. In that sense, an  unbiased forecaster is a  theoretical construct. As a side comment, we note that the judgment of ordinary people is biased in several ways. For instance, there is the so-called anchor effect,  in which mentioning in passing  a figure, say 400, to a group of  randomly picked individuals  who are asked to guess the number of candies in a jar and a different figure, say 800, to another group exposed to the same jar leads to very distinct collective predictions \cite{Sunstein_06}. Accordingly, in \ref{AppB} we analyze the candies-in-a-jar experiment \cite{Davi_20}
 and show  that  the resulting  collective estimate obtained from the aggregation of laypeople estimates cannot be explained by the  unbiased estimates assumption.

\section{The FRBP forecast database}\label{sec:data} 

\begin{figure}  
\centering  
 \includegraphics[width=.47\textwidth]{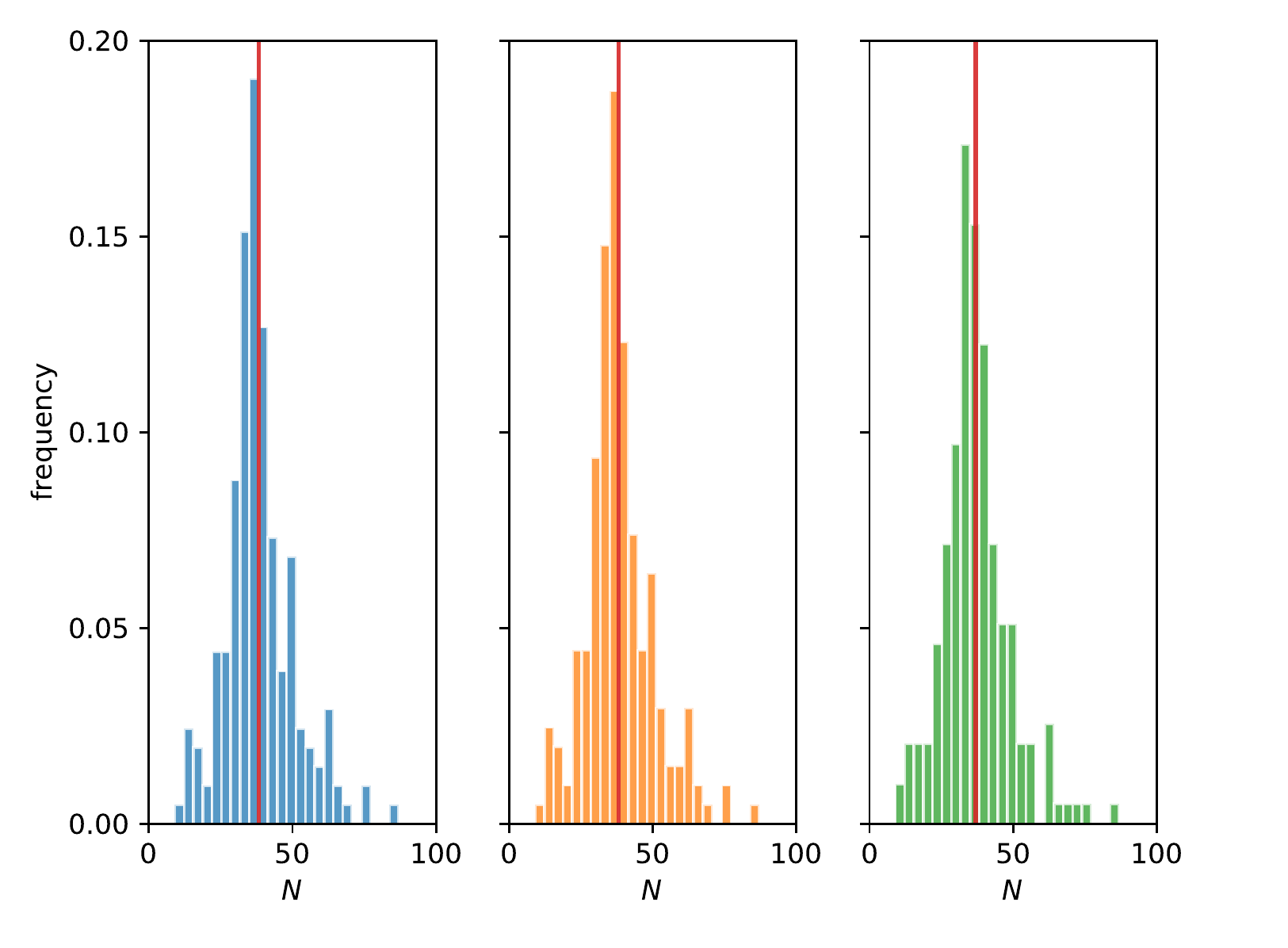}  
 \caption{Histograms of the number of participants $N$ in the 205  short-range  forecasts  (left panel), in the 203 medium-range forecasts (middle panel) and in the 196 long-range forecasts  (right panel) on the nominal gross domestic product of the Federal Reserve Bank of Philadelphia's Survey of Professional Forecasters. The vertical  red lines indicate  the average number of participants in each forecast range. }  
\label{fig:1}  
\end{figure}

The Federal Reserve Bank of Philadelphia's (FRBP) Survey of Professional Forecasters  offers quarterly projections for five quarters of a variety of economic indicators \cite{FRBP}. In order to lighten the analysis, here we will focus mainly on  semestrial forecasts of the nominal gross domestic product (NGDP), as our  conclusions apply to the other indicators as well. In particular, we consider the forecast for the
current quarter, that is, the quarter when the   survey was conducted and  the forecasts for two and four quarters  later. Henceforth we will refer to these forecasts as  short-range, medium-range  and long-range forecasts.  
We use the NGDP  forecasts  available in the FRBP database  from the fourth quarter of 1968 to the fourth quarter of 2019. 
 There are $1 + (2019-1969 +1)\times 4=205$ quarters in total which equals the number of short-range  estimate distributions. In addition, since the medium-range forecasts predict  the economic indicators  two quarters  after the quarter when the survey was conducted, the last two quarters of 2019 are not used for  those forecasts  and so the number of medium-range  estimate distributions is 203. Following this reasoning, the number of long-range  estimate distributions should be 201. However,  the  data for the long-range forecasts are missing in the first, second and third quarters of 1969, in the first quarter of 1970 and in the third quarter of 1974, so that the number of long-range  estimate distributions  available in the FRBP database is 196. 
All forecasters are select  economists and  the mean number of   forecasters in each  experiment  is about  37.  Figure \ref{fig:1} shows the histograms of the number of participants $N$ in each experiment for the three forecast ranges. The minimum number of participants is $N=9$ and the maximum is $N=87$ in any forecast range.
 
The  separation of the experiments on short-range, medium-range and long-range  forecasts allows the control of the difficulty of the  forecasts and the  study of its influence   on the distribution of the individual estimates.  We note that in Ref. \cite{Nash_14} these distinct  forecast ranges were  clumped together to form a single large forecast database on the NGDP. We think, however, that this procedure is not appropriate as the range of the  forecasts  has a major effect on the distribution of the estimates, as we will see next.

\section{Results}\label{sec:res}

\begin{figure}  
\centering  
 \includegraphics[width=.47\textwidth]{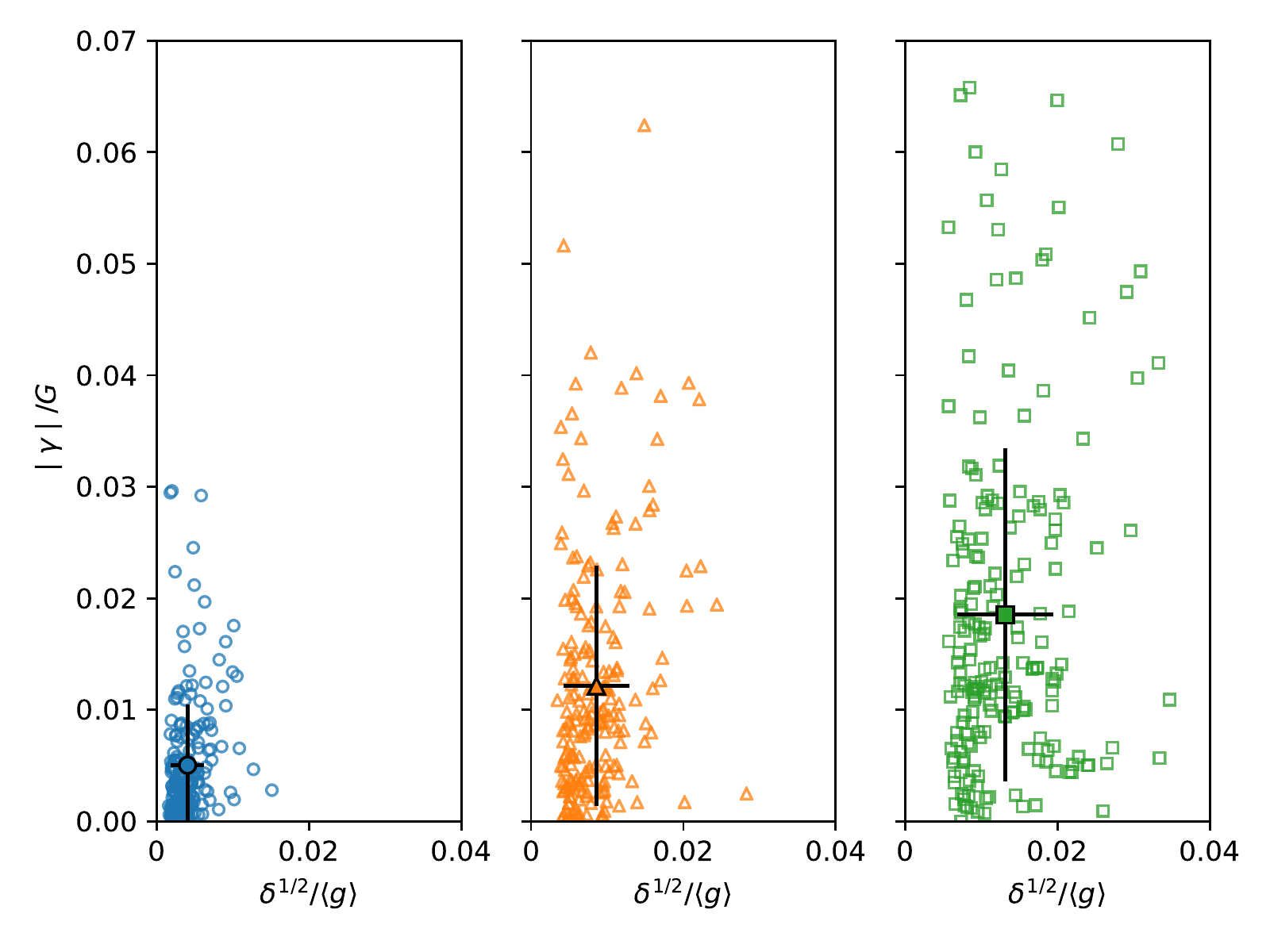}  
 \caption{Scatter plots of  the scaled diversity $\delta^{1/2}/\langle g \rangle $ and the scaled collective error $| \gamma |/G$ for the  short-range  (left panel), medium-range  (middle panel)  and long-range  (right panel) NGDP  forecasts.  The filled symbols and the  horizontal and vertical  lines indicate the means and  the  standard deviations.}  
\label{fig:2}  
\end{figure}

\begin{figure}
\centering  
 \includegraphics[width=.47\textwidth]{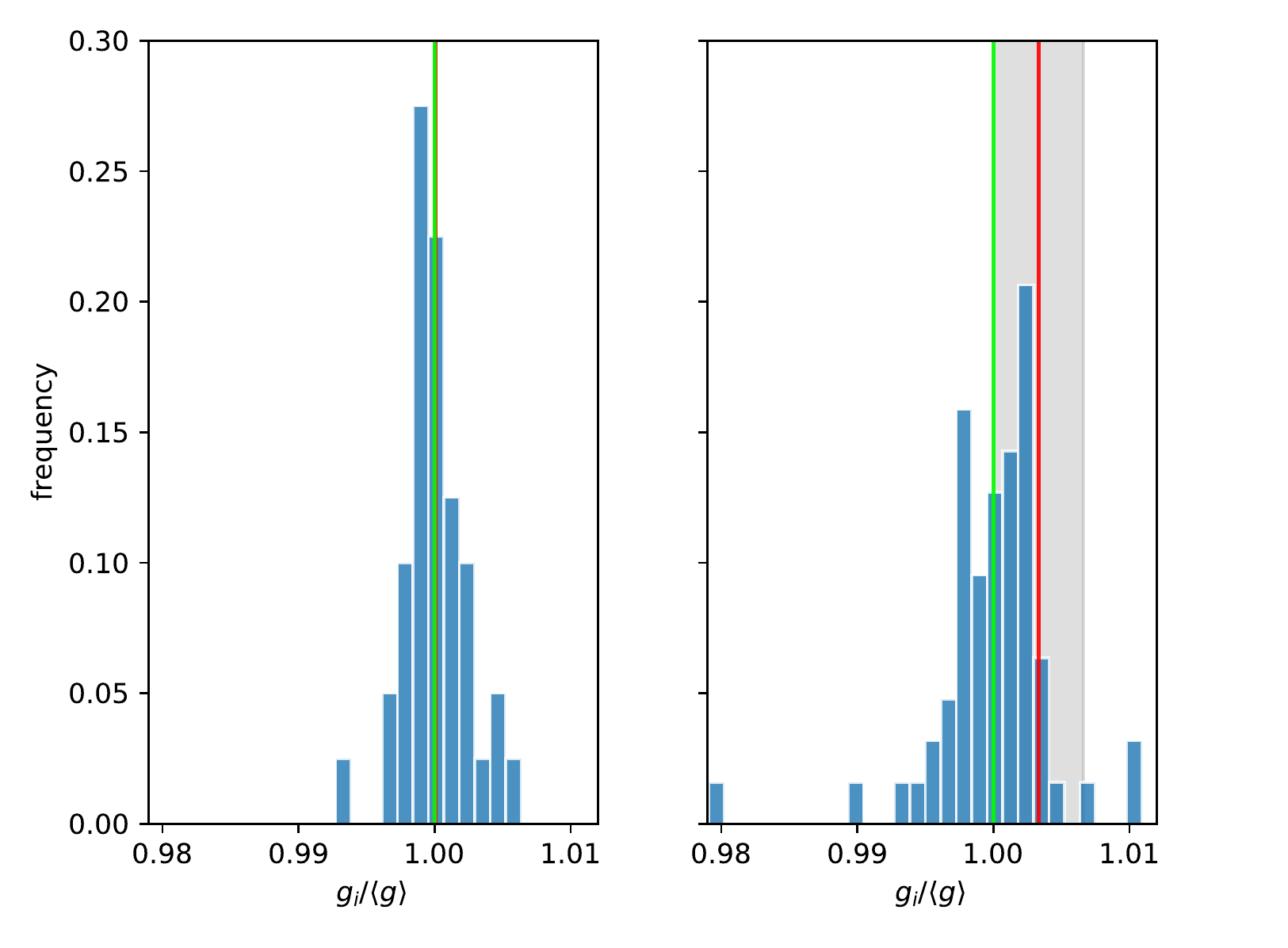}  
 \caption{Histograms of the relative estimates $g_i/ \langle g \rangle$ for two short-range forecasts. The vertical red lines indicate the ratio between the true value of the NGDP indicator and the crowd estimate, i.e., $G/\langle g \rangle$, whereas the vertical green lines at 
 $g_i/ \langle g \rangle = 1$ indicate the scaled  crowd estimate.  In the left panel there are $N=40$ participants and none of them predicted better than the crowd, whereas in the right panel 
 there are $N=63$ participants and  43\% of them predicted more accurately than the crowd. The gray region highlights the individual estimates that are better than the crowd's.}  
\label{fig:3}  
\end{figure}

We begin our analysis with the study of the correlation between the scaled crowd accuracy $| \gamma |/G$ and the scaled diversity of the estimates $\delta^{1/2}/\langle g \rangle$. Figure \ref{fig:2} shows the scatter plots of these quantities  for the three different forecast ranges, where each data point represents a particular forecast experiment.  As already pointed out, since  the number of  participants  $N$ as well as the NGDP true value $G$ may vary in different experiments for the same forecast range (see, e.g.,  figure \ref{fig:1}), it is  necessary to  consider dimensionless summary statistics to properly compare the outcomes of the experiments.  We recall that each experiment produces a  distribution of estimates from where we extract the relevant summary statistics.  For instance, figure \ref{fig:3}  illustrates two  distributions of estimates for the short-range forecast scenario, which correspond to two different data points in the left panel of figure \ref{fig:2}. The individual estimates in the x-axis of figure \ref{fig:3}  are scaled by the average estimate $\langle g \rangle $ defined in equation (\ref{av}), which is by definition the crowd prediction, so that the scaled crowd prediction  is  $\langle g \rangle / \langle g \rangle = 1$.  Since the  scaled true value is  $G / \langle g \rangle $,   the participants that predict better than the crowd are those whose estimates   satisfy $\mid g_i/\langle g \rangle - G/\langle g \rangle \mid <  \mid 1 - G/\langle g \rangle \mid$, i.e, are those whose individual errors are smaller than the error of the crowd.  

Figure \ref{fig:2} shows that the accuracy $| \gamma |/G$ and the dispersion  $\delta^{1/2}/\langle g \rangle$ of the estimates vary considerably as the forecast range increases, indicating that  clumping  those forecasts together as done in Ref.  \cite{Nash_14} may not be a judicious choice. In particular,    the short-range forecasts are on the average about three times more accurate and  four times  less disperse than the long-range forecasts. This is somewhat expected since the farther the forecast range, the greater the odds that the predicted  indicator will be influenced by  unforeseen factors. 

Since the two summary statistics displayed in the scatter plots of figure \ref{fig:2} (as well as in all   scatter plots exhibited in this paper) are probably not linearly dependent, we use  the nonparametric Spearman correlation coefficient $\rho$ to measure correlations rather than the usual  Pearson's coefficient, which is designed to quantify linear correlations.   We find, however, that  these two correlation coefficients yield  very similar values for the quantities considered here. In addition, the statistical significance of the measured correlation coefficients is determined by their $p$-values that yield the probability  that the same coefficients are obtained if the null hypothesis is true, i.e., if  the two summary statistics are uncorrelated.  Hence,  a very small $p$-value means that the observed  correlation is very unlikely under the null hypothesis and so that it is  statistically different from zero.

The Spearman correlation coefficient between  $\delta^{1/2}/\langle g \rangle$ and  $| \gamma |/G$  is  $\rho = 0.31$ with $p$-value  $ < 10^{-6}$  for the short-range, $\rho = 0.25$ with  $p$-value $  < 10^{-6}$  for the medium-range  and $\rho = 0.14$ with  $p$-value $= 0.05$ for the long-range forecasts.  The unfounded interpretation of the diversity prediction theorem  that associates a high prediction diversity  to a low collective error implies a  negative correlation between $\delta^{1/2}/\langle g \rangle $ and $| \gamma |/G$, which is clearly  not supported by our findings. In fact, given that the forecasters are all expert economists, it is somewhat intuitive  to expect that the less disperse  their estimates, the closer they are to the true value.  For the long-range forecast, their expertise becomes less influential to the success of the predictions and so the (statistically significant)  positive correlation between collective error and diversity decreases \cite{Mauboussin_12}. We note that this positive correlation  is a prediction of the augmented quincunx model \cite{Nash_14} and that it holds also for the unbiased estimates assumption,   as shown in \ref{AppA}. The reason a positive correlation between the diversity (or variance) of the estimates and the collective error is consistent with the diversity prediction theorem is because  the average quadratic individual error increases together with  the variance.

\begin{figure} 
\centering  
 \includegraphics[width=.47\textwidth]{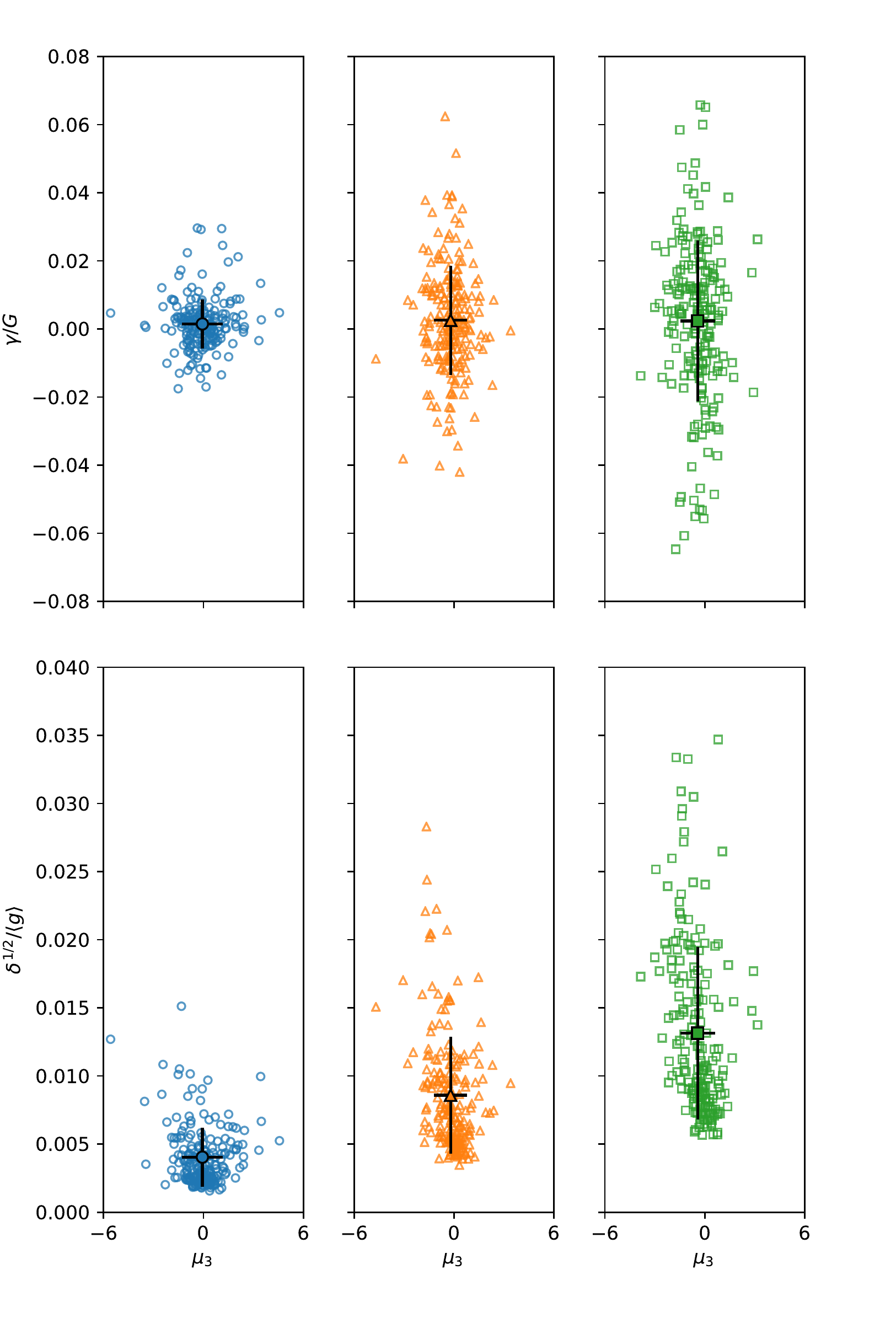}  
 \caption{Scatter plots of  the skewness $ \mu_3  $  and  the scaled collective error $ \gamma/G$ (upper set of panels) and  of $ \mu_3  $ and the scaled diversity $\delta^{1/2}/\langle g \rangle $ (lower set of  panels) for short-range  (left panels), medium-range  (middle panels)  and long-range  (right panels) NGDP  forecasts.  The filled symbols and the  horizontal and vertical  lines indicate the means and  the  standard deviations.}  
\label{fig:4}  
\end{figure}

An interesting  outcome of the augmented quincunx model refers to the role of the skewness $\mu_3$  of the distribution of estimates. In particular, that model predicts  a negative correlation between $\mu_3$  and the (signed) collective error $\gamma$.  Accordingly, in figure \ref{fig:4}  we show scatter plots of $\mu_3$ and $ \gamma /G$  as well as of $\mu_3$ and $\delta^{1/2}/\langle g \rangle$  for the different forecast ranges. The Spearman correlation coefficient between  $ \mu_3 $   and  $ \gamma /G$  is  $\rho = 0.008$ with   $p$-value $= 0.91$ for the short-range,  $\rho = -0.03$  with $p$-value $= 0.62$  for the medium-range  and $\rho = -0.12$  with $p$-value $= 0.09$ for the long-range forecasts.  The low values of these coefficients and their high $p$-values   point to the   
little  relevance of the skewness of the estimate distributions to the crowd prediction,  in  disagreement with the claims of Ref.\ \cite{Nash_14}. In addition, our results indicate that  the skewness is weakly affected by the forecast range. This  finding contrasts with  the results for the  dispersion of the estimates that  is strongly affected by that range.  In fact, our results suggest that the more difficult the forecasts, in the sense that there is more room for noise to alter the future outcome,  the  greater  the variance  of the  estimates. The   Spearman correlation coefficient between  $ \mu_3 $ and   $\delta^{1/2}/\langle g \rangle $  is  $\rho = -0.09$ with $p$-value $= 0.19$ for the short-range,  $\rho= -0.36$ with  $p$-value  $ < 10^{-6}$ for the medium-range and  $\rho = -0.46$ with  $p$-value  $ < 10^{-6}$
 for the long-range forecasts, which implies that, at least for the medium and long-range forecasts,  a large variance is associated with  left-tailed  estimate distributions.  This interesting and statistically significant correlation is not predicted by the  augmented quincunx model since   the sign of the skewness is not affected by the variance  in that model \cite{Nash_14}. In addition, we note that the unbiased estimates assumption predicts no significant  correlation between $\mu_3$ and $\delta^{1/2}/\langle g \rangle$ (see \ref{AppA}), so the negative correlation between these quantities reported here is a genuine effect of the aggregation of    real-world biased forecasts.

\begin{figure}
\centering  
 \includegraphics[width=.47\textwidth]{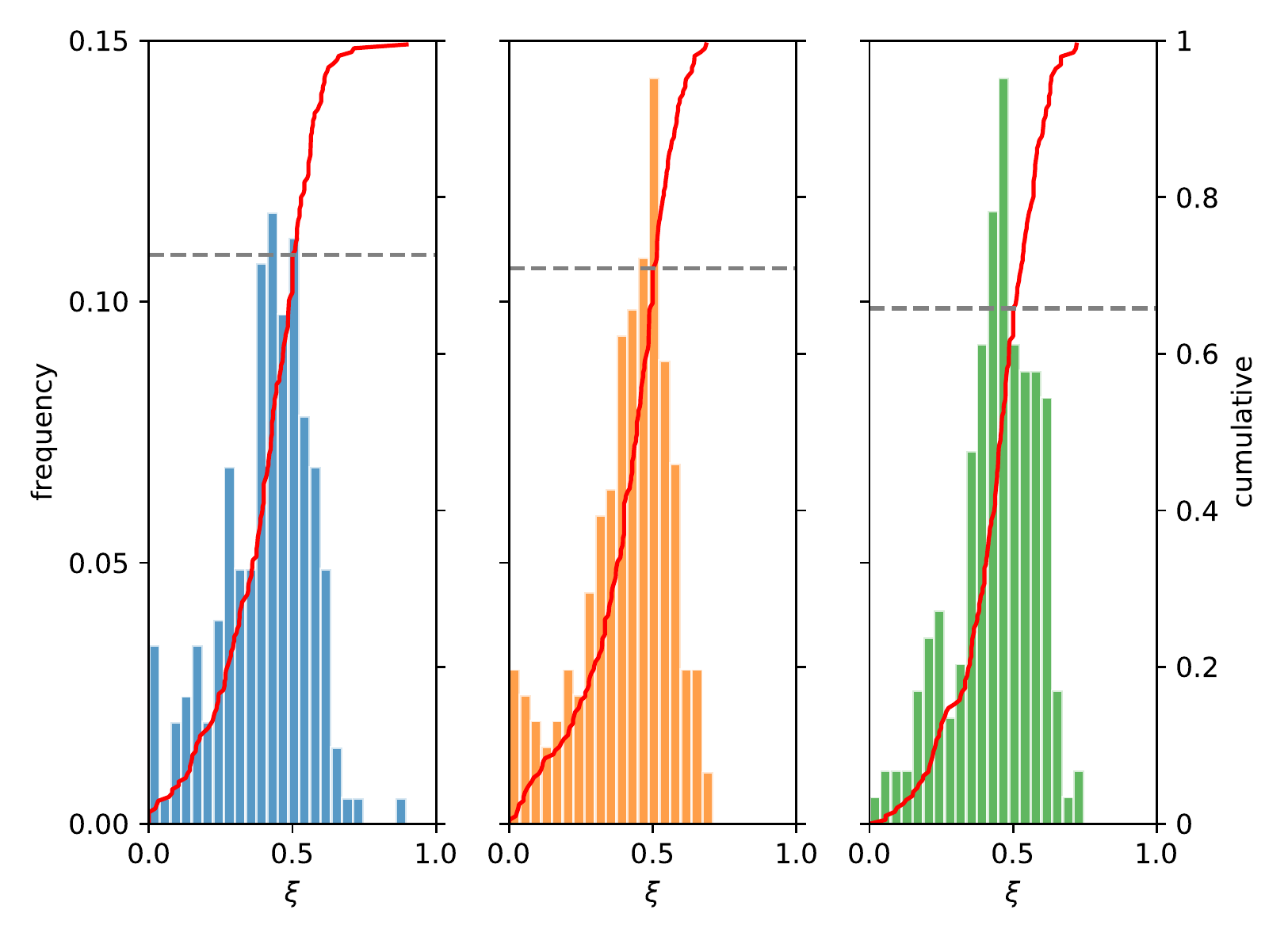}  
 \caption{Histograms of the proportion of experiments for which  a fraction $\xi$ of the individual estimates  are more  accurate than the collective estimate for   short-range  (left panel), medium-range  (middle panel)  and long-range  (right panel) forecasts. The red curves are the cumulative distributions and the  horizontal dashed  lines indicate the values of the cumulative distributions at $\xi = 0.5$. }  
\label{fig:5}  
\end{figure}

We conclude our analysis by challenging the common view that  the crowd is superior to  most  of its integrants \cite{Surowiecki_04}.   In fact, if people believed that  an individual selected  at random  had a fair chance of beating the crowd, the idea of the wisdom of crowds  would probably never have taken off. Here we  address this issue quantitatively by   measuring   the fraction  of individual estimates that are superior to the collective estimate for each forecast experiment. The results are presented in form of histograms and cumulative distributions in figure \ref{fig:5}, where  the height of the bars is the proportion of experiments for  which that fraction equals $\xi \in \left [ 0, 1 \right ]$. There are a few experiments with $\xi =0$ so that the crowd beats all    individuals  and  one short-range forecast experiment  where 85\% of the individuals beat the crowd. The fraction of experiments for which the crowd  is more accurate than the majority of the participants is $150/205 \approx 0.73$ for the short-range forecasts, $145/203 \approx 0.71$ for the medium-range forecasts and $130/196 \approx 0.66$ for the long-range forecasts. These figures are obtained by evaluating the cumulative distribution function at $\xi=0.5$.  Hence,  a randomly chosen individual  has  probability greater than  $1/2$ of beating the crowd in about 30\% of the experiments reported here. 

We stress that finding that   85\% of the individuals beat the crowd in a particular  forecast experiment does not contradict Page's diversity prediction theorem, which asserts that the collective error is always less than the average individual error, i.e., $| \gamma | \leq \epsilon^{1/2}$. In fact,  that  particular experiment involved  $N=39$ participants among which 4 outliers  produced completely off the mark estimates resulting in the  inflation of the mean individual error. This point betrays the fact that the theorem (\ref{DPT}) is largely irrelevant for practical issues concerning the use or not of the crowd as an effective forecaster.

\begin{figure}
\centering  
 \includegraphics[width=.47\textwidth]{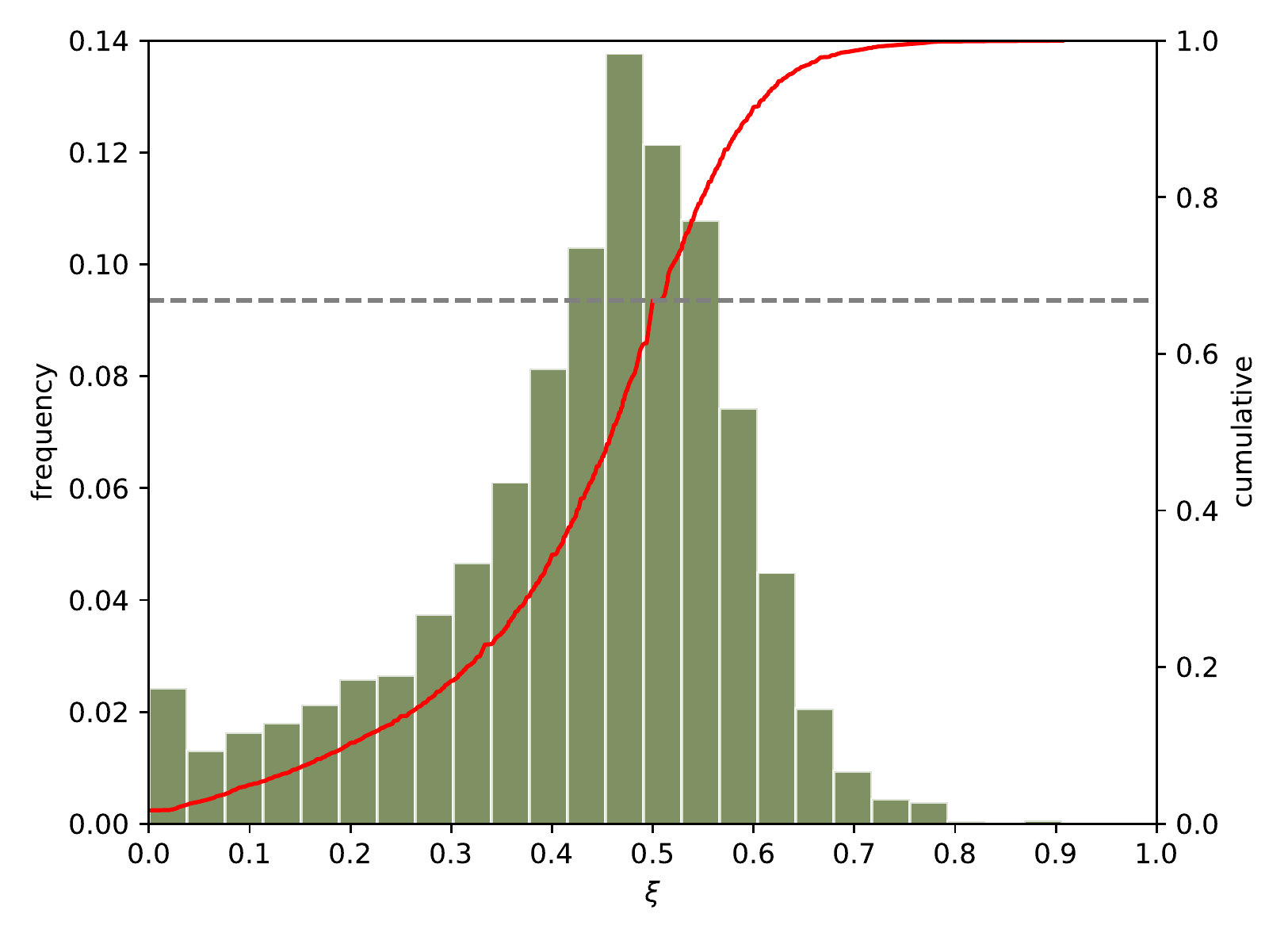}  
 \caption{Histograms of the proportion of experiments for which  a fraction $\xi$ of the individual estimates  are more  accurate than the collective estimate.  The  data comprises the $8650$ experiments of the FRBP forecast database for  ten distinct economic indicators and five forecast ranges. The red curve is the cumulative distribution and the  horizontal dashed  line indicates the value of the cumulative distribution at $\xi = 0.5$. }  
\label{fig:6}  
\end{figure}

To support these findings, in  figure \ref{fig:6} we clump together  $8650$ experiments of the FRBP forecast database  without regard to 
 the economic indicator (there are ten distinct indicators) or to  the range of the forecast (there are five distinct ranges).  We find that the crowd is superior to any individual in only  $1.7 \%$ of the experiments (viz., those for which $\xi=0$), whereas it is superior to most individuals in $66.8 \%$ of the experiments (viz., those for which $\xi \leq 1/2$).  Hence the widespread claims about the superiority of the crowd over  the individuals \cite{Surowiecki_04} is most likely an artifice of selective attention that gives prominence to  successful  outcomes only. Those claims are legitimate only for unrealistic crowds composed of unbiased forecasters, as shown in \ref{AppA}.

 \section{Discussion}\label{sec:disc}

It  is almost a clich\'e   to remark that a  group of cooperating individuals can solve problems more efficiently than when those individuals work in isolation  \cite{Huberman_90,Clearwater_91}.   Cooperation is, in general, a successful problem solving strategy \cite{Bloom_01}, though it is not clear whether it merely speeds up the time to find the solutions, or whether it alters qualitatively the statistical signature of the search for the solutions \cite{Reia_19,Reia_20}. Yet, in some cases, cooperation may well lead the group astray resulting in the madness of crowds \cite{MacKay_41} or, less dramatically,  it may simply undermine  the benefits of combining independent  forecasts \cite{King_11,Lorenz_11}.
 
A  rather peculiar manner to circumvent the potential negative effects of cooperation while still benefiting  from the group intelligence is the so-called wisdom of crowds, i.e., the notion that  a  collection of independently deciding individuals is likely to predict better than individuals or even experts within the group  \cite{Surowiecki_04}, which ironically seems to  have become itself a piece of crowd wisdom \cite{Prelec_17}. The first report of this phenomenon  in the literature was probably  Galton's  account of the surprisingly accurate estimate of the  weight of an ox  given by the median of the sample of the individual guesses \cite{Galton_07}.
 
Although much of the evidence of the wisdom of crowds  is anecdotal (see, e.g., \cite{Surowiecki_04,Sunstein_06}), there are a few efforts aiming at  explaining this phenomenon  either using a purely statistical  rationale \cite{Armstrong_01,Page_07} or using  psychophysical arguments on the nature of the individual estimates \cite{Nash_14}. Typically,   these approaches aim at  inferring  the quality of the crowd estimate using information about the distribution of individual estimates (see, e.g., figure \ref{fig:3}), such as the variance and the skew of the estimates.
Here we address the soundness of those explanations using  forecasts of economic indicators from the Federal Reserve Bank of Philadelphia's (FRBP)  Survey of Professional Forecasters  database \cite{FRBP}. The difficulty of the forecasts can be tuned by controlling for the forecast range.

A word is in order about the particularity that the FRBP database  comprises the predictions of expert economists, whereas the common   view of the wisdom of crowds is that it comprises the predictions of ordinary people.  In fact,  the wisdom of crowds  as a method of information aggregation depends on the presence of an expressive number of experts in the crowd since nothing good can come from  the aggregation of random information: if every person in France were asked to specify the gross domestic product of India, the average answer would probably be wildly off \cite{Sunstein_06}.  The widespread interest   in the wisdom of crowds stems from the possibility of combining the forecasts of experts  in the hope that  many expert minds are   better than a few. As a matter of fact, just after the publication of  Galton's paper on the ox-weighing experiment  \cite{Galton_07}, it was pointed out that Galton had not been exposed to \textit{Vox Populi} but to \textit{Vox Expertorum},   as  the participants of the contest were butchers and farmers whose livelihood depended on their ability to judge the weight of farm animals before trading \cite{Coste_07}. Moreover, the 6 pence    tickets probably deterred the participation of dilettantes in that celebrated contest.

We stress that the main focus  of the research  on information aggregation is on the prediction of economic, political and  other valuable indicators  so that the forecasters are  necessarily experts since  there is no point in  asking ordinary people to predict,   say, the NGDP at the end of the current year. Nevertheless, in \ref{AppB} we revisit a few  wisdom-of-crowds experiments where the participants are  laypeople \cite{Davi_20}.   The experiments are  the estimates of the number of candies in a jar, the length of a paper strip, the number of pages of a book and the weight of a bag of beans. These are the kind of factual  questions that make sense asking ordinary people.   The results generally agree with our findings using the  FRBP database. In particular, the probability that a randomly picked participant outperforms the crowd varies from $15\%$ for the paper-strip experiment to $38\%$ for the pages-of-a-book experiment, which support our main point that there is a fair chance that  random participant will beat the crowd.

Our results suggest that the difficulty of the forecasts is associated with  large   variances and with long   left tails of the distributions of estimates. In addition, we find that a large variance is associated with a poor crowd accuracy, in disagreement with the interpretation of the  diversity prediction theorem that  the increase of the diversity of the estimates  leads to a decrease of the   collective error \cite{Page_07}. Moreover,  when controlling for the forecast range, we find no  evidence of an association between the  skew of the estimates  and the collective error, in disagreement with the predictions of the augmented quincunx model \cite{Nash_14}. 

We pay special attention to the unbiased estimates assumption that explains the accuracy of the crowd by conjecturing that  the errors of the individual estimates spread in equal proportion around the true value of the unknown quantity so that they cancel out when those estimates are  combined together \cite{Bates_69}.  In \ref{AppA} we replicate the forecast experiments by replacing the economists by (virtual) unbiased forecasters and show that much of the hailed  features of the wisdom of crowds are properties of the combination of  unbiased forecasts instead.  In that appendix, we also offer a statistical test to check if the economists' forecasts  in the FRBP database are in fact biased  and  find that  $85\%$ of the collective predictions   have $p$-value less than $0.05$, which indicates that they are unlikely to be obtained by the aggregation of unbiased forecasts. In addition, in \ref{AppB} we show that the unbiased estimates assumption  cannot be rejected in the case the task is such that  the participants are likely to deliver accurate estimates, as in the paper-strip experiment.  However, it can safely be rejected in the case that the participants deliver far off estimates as in the  candies-in-a-jar experiment.  These findings support the view that unbiased forecasters are experts.

The wisdom of real crowds is very different from the apocryphal wisdom of crowds of unbiased forecasters. For instance,  the crowd beats all individuals in only  around  2\% of the FRBP  forecast experiments and it  beats most individuals  in about  70\% of those experiments, whereas the corresponding figures for the unbiased forecasters are about 16\% and 100\%, respectively. Hence, since  there is a fair chance that a randomly chosen individual will beat the crowd in real-world forecasts, the only explanation that we can find for the popularity of the wisdom of crowds is selective attention (or cherry picking) that gives prominence to  outcomes that tally with Galton's conclusions.

\bigskip

\acknowledgments
The research of JFF was  supported in part 
 by Grant No.\  2020/03041-3, Fun\-da\-\c{c}\~ao de Amparo \`a Pesquisa do Estado de S\~ao Paulo 
(FAPESP) and  by Grant No.\ 305058/2017-7, Conselho Nacional de Desenvolvimento 
Cient\'{\i}\-fi\-co e Tecnol\'ogico (CNPq).
SMR  was supported by the Coordena\c{c}\~ao de Aperfei\c{c}oamento de Pessoal de N\'{\i}vel Superior - Brasil (CAPES) - Finance Code 001.

\section*{Appendix A}\label{AppA}

\renewcommand{\thefigure}{A\arabic{figure}}
\renewcommand{\theequation}{A\arabic{equation}}
\setcounter{figure}{0}
\setcounter{equation}{0}

\begin{figure}  
\centering  
 \includegraphics[width=.47\textwidth]{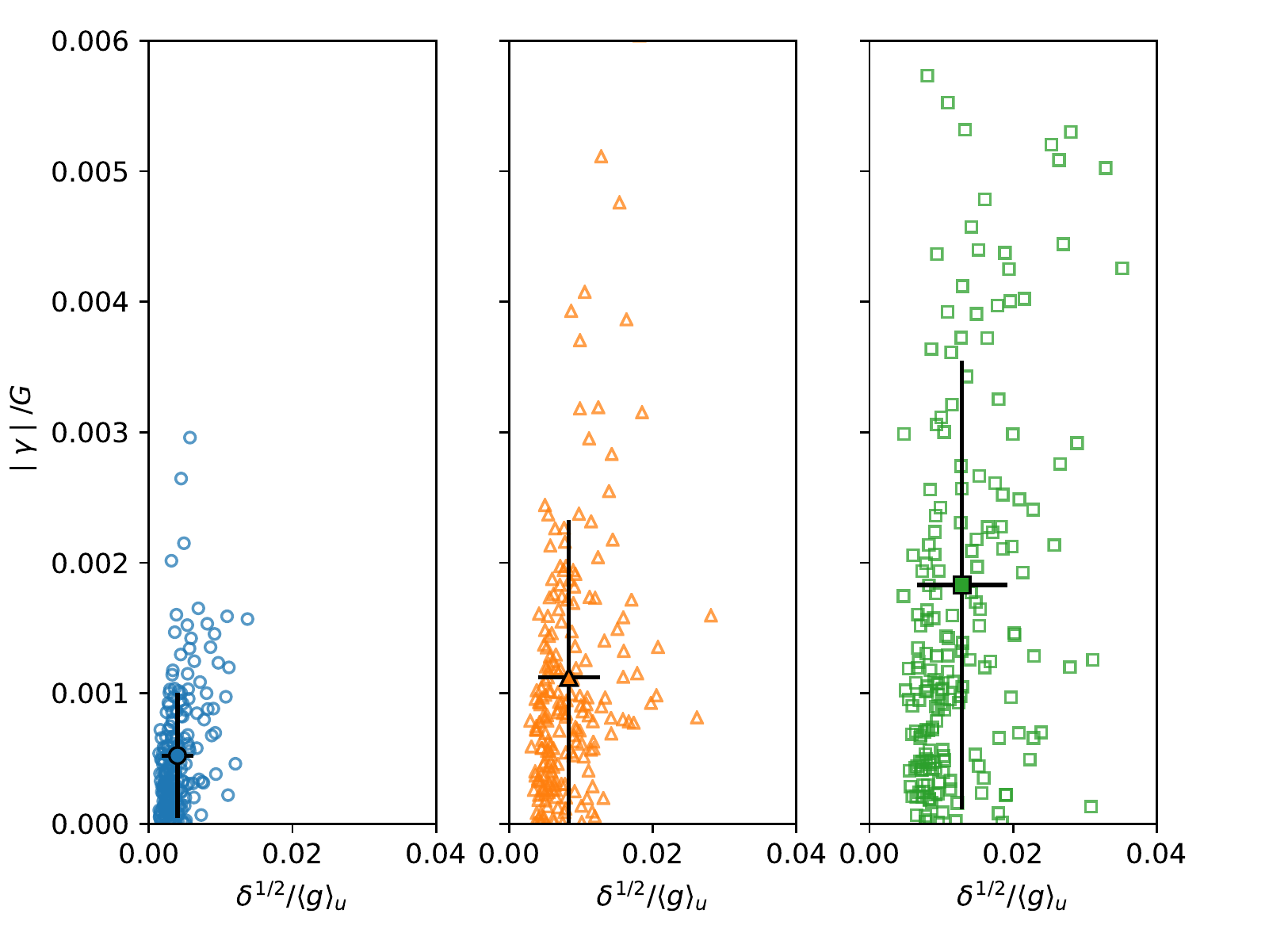}  
 \caption{ Scatter plots of the scaled diversity  $\delta^{1/2}/\langle g \rangle $ and the  relative collective error $| \gamma |/G$ for short-range  (left panel), medium-range  (middle panel)  and long-range  (right panel)  unbiased forecasts. The filled symbols and the  horizontal and vertical  lines indicate the means and  the  standard deviations. }  
\label{fig:A1}  
\end{figure}

In this appendix we  examine  the predictions of the  unbiased estimates assumption  for  the forecast experiments considered in the main text. In  particular, for each experiment with a given number $N$ of participants, we calculate the variance of the estimates $\delta$ and use it to produce $N$ independent  unbiased estimates distributed according to a Gaussian  of mean $G$ (the true value of the economic indicator) and variance $\delta$, i.e.,
\begin{equation}\label{uea}
P_{u} \left ( g_i \right ) = \frac{1}{\sqrt{2 \pi \delta}} \exp \left [ - \frac{ \left ( g_i - G \right )^2}{2 \delta} \right ]
\end{equation}
for $i=1, \ldots, N$. Hence, by construction, the diversity of the virtual unbiased estimates equals the diversity of the economists' forecasts. Although we split the data in three forecast ranges, as done in the main text, we stress that from the perspective of the unbiased forecasters  the only difference between those ranges is the variance $\delta$ of the estimates. In addition, we note that for  the   values of $N$ considered here (see, e.g., figure \ref{fig:1}), the arithmetic average  $\langle g \rangle_u  = \sum_i^N g_i/N$   is expected to differ  from $G$. In fact, the prediction of the unbiased crowd $\langle g \rangle_u$  is a random variable distributed according to the Gaussian
\begin{equation}\label{up}
P  \left ( \langle g \rangle_u  \right ) = \frac{1}{\sqrt{2 \pi \delta/N}} \exp \left [ - \frac{ \left ( \langle g \rangle_u - G \right )^2}{2 \delta/N} \right ]
\end{equation}
since the individual estimates $g_i$ are statistically independent random variables.

Figure \ref{fig:A1} shows the scatter plots of  $\delta^{1/2}/\langle g \rangle_u $ and $| \gamma |/G$ for the unbiased forecasters. As expected, the crowd estimate in this case  is about ten times more accurate than in the original forecast  experiments. The Spearman correlation coefficient between  $\delta^{1/2}/\langle g \rangle_u$ and  $| \gamma |/G$  is  $\rho = 0.39$ with  $p$-value  $ < 10^{-6}$ for the short-range, $\rho = 0.37$ with  $p$-value  $ < 10^{-6}$  for the medium-range  and $\rho = 0.44$ with  $p$-value  $ < 10^{-6}$  for the long-range forecasts. Hence, the positive and statistically significant  correlation between the  diversity of the estimates and the crowd accuracy is more pronounced for the unbiased forecasts  than for the economists forecasts.

\begin{figure}  
\centering  
 \includegraphics[width=.47\textwidth]{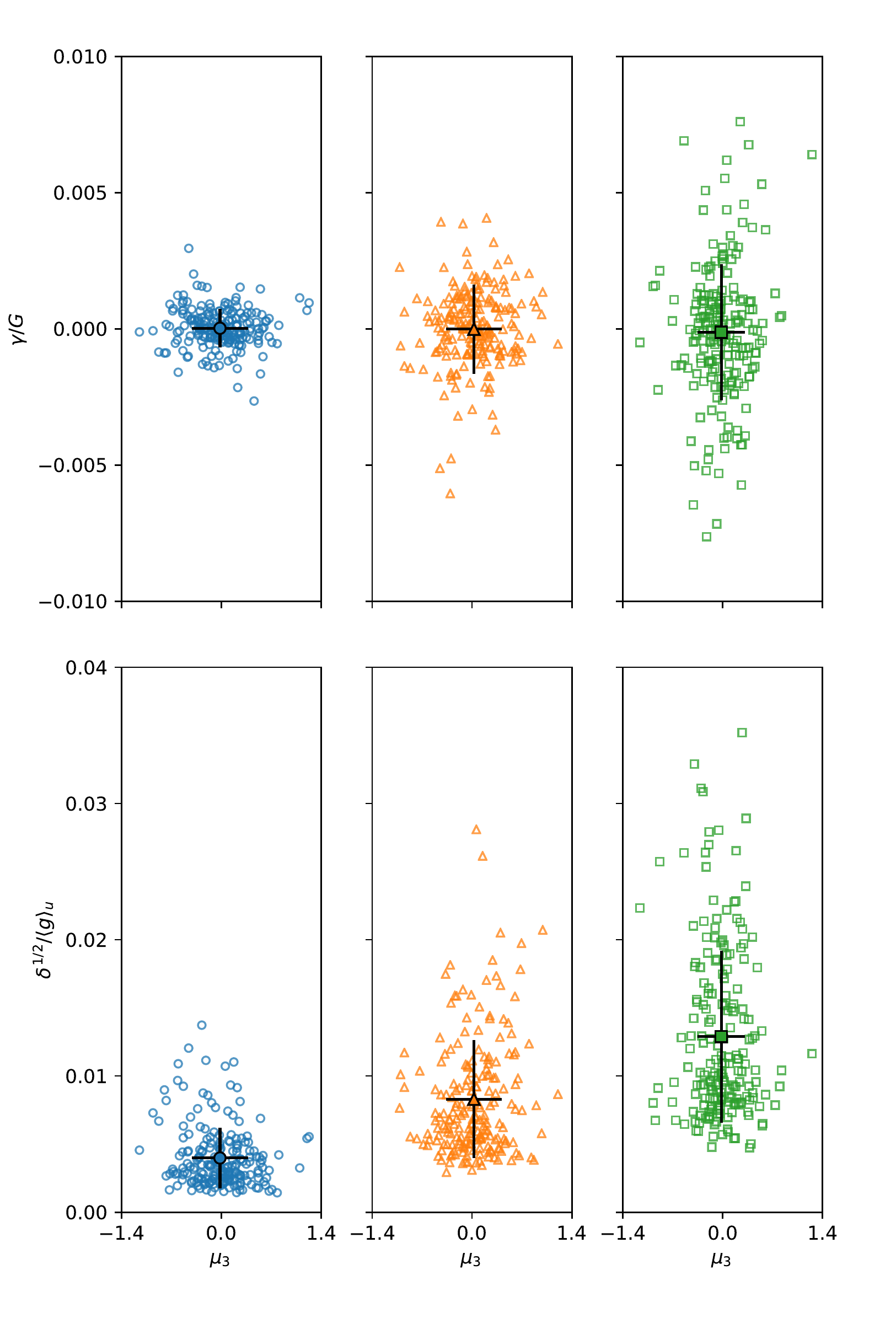}  
 \caption{Scatter plots of  the skewness $ \mu_3  $  and  the scaled collective error $ \gamma/G$ (upper set of panels) and  of $ \mu_3  $ and the scaled diversity $\delta^{1/2}/\langle g \rangle_u $ (lower set of  panels) for short-range  (left panels), medium-range  (middle panels)  and long-range  (right panels) unbiased  forecasts.  The filled symbols and the  horizontal and vertical  lines indicate the means and  the  standard deviations. }  
\label{fig:A2}  
\end{figure}

Figure \ref{fig:A2} shows the scatter plots of  $\mu_3 $ and $ \gamma /G$  as well as of  $\mu_3 $ and $\delta^{1/2}/\langle g \rangle_u $ for the unbiased forecasters. The noteworthy aspect here is the small range of variation  of the skewness values as compared to the results of figure \ref{fig:4}. Of course, the nonzero values of $\mu_3$  are due to the small number of estimates $N$ in each experiment, since the expected skewness of a Gaussian is zero.  The Spearman correlation coefficient between  $\mu_3$ and $ \gamma /G$  is  $\rho = -0.09$  with $p$-value $= 0.20$ for the short-range, $\rho = 0.06$ with $p$-value $= 0.35$  for the medium-range  and $\rho = 0.02$ with $p$-value $= 0.73$ for the long-range forecasts. These coefficients are similar to those of the real experiments and their high $p$-values indicate that the skewness of the estimates offers no information on the  crowd prediction, regardless of the nature of the forecasters.
An unexpected result of our analysis of the economists' forecasts is the  negative correlation between the diversity and the skewness of the estimates (see lower set of panels in figure \ref{fig:4}), which implies that a large variance is associated with a long left tail of the distribution of estimates. The results of the  lower set of panels in figure \ref{fig:A2} point to a different conclusion. In fact,  the Spearman correlation coefficient between  these quantities  is  $\rho = -0.11$ with  $p$-value $= 0.10$ for the short-range, $\rho = 0.12$ with  $p$-value $= 0.09$ for the medium-range  and $\rho = -0.03$ with $p$-value $= 0.64$  for the long-range forecasts. Again, the low values and  the inconsistency of the signs  of these coefficients as well as their high $p$-values   suggest that the skewness plays no role at all on the outcome of unbiased forecasts.


\begin{figure}
\centering  
 \includegraphics[width=.47\textwidth]{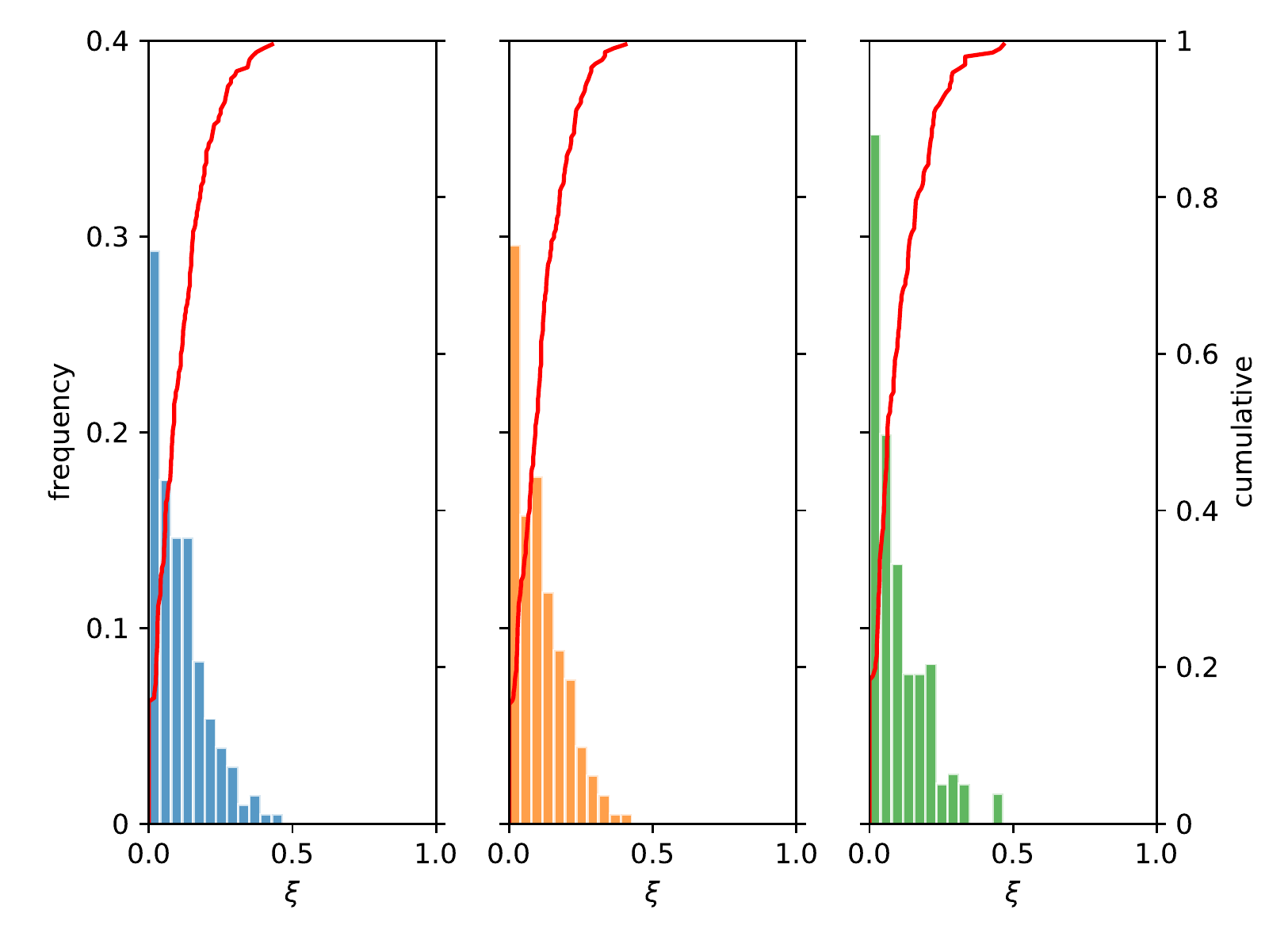}  
 \caption{Histograms of the proportion of experiments for which  a fraction $\xi$ of the individual estimates  are more  accurate than the collective estimate for   short-range  (left panel), medium-range  (middle panel)  and long-range  (right panel) unbiased forecasts.  The red curves are the cumulative distributions. }  
\label{fig:A3}  
\end{figure}

\begin{figure}
\centering  
 \includegraphics[width=.47\textwidth]{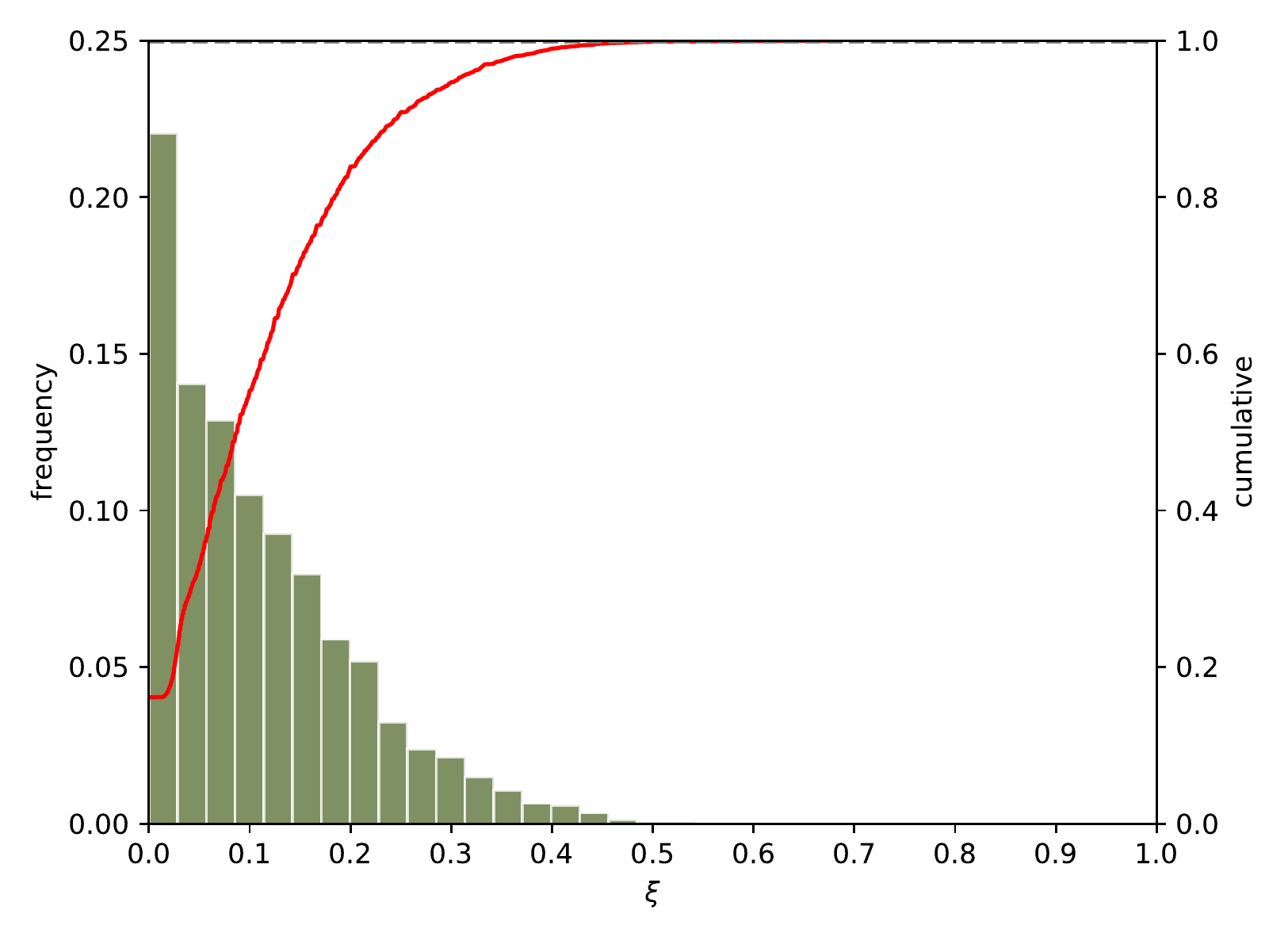}  
 \caption{Histograms of the proportion of experiments for which  a fraction $\xi$ of the individual unbiased estimates  are more  accurate than the collective estimate.  The  expert forecasts of the  $8650$ experiments of the FRBP   database were  replaced by virtual unbiased forecasters. The red curve is the cumulative distribution. }  
\label{fig:A4}  
\end{figure}

Although the previous scatter plots show only mild quantitative differences between the economists and the unbiased forecasters, the advantage conferred to  the crowd over its  members differs starkly between  these two types of forecasters.  Figure \ref{fig:A3}, which  shows the  histograms and the cumulative distributions of  the number of experiments for which a fraction $\xi$ of individuals beat the crowd, illustrates this point. In fact, the most probable outcome is that the crowd beats all individuals ($\xi = 0$) in the case of unbiased forecasters, in contrasts to our findings for the human experts (figure  \ref{fig:5}).

\begin{figure}
\centering  
 \includegraphics[width=.47\textwidth]{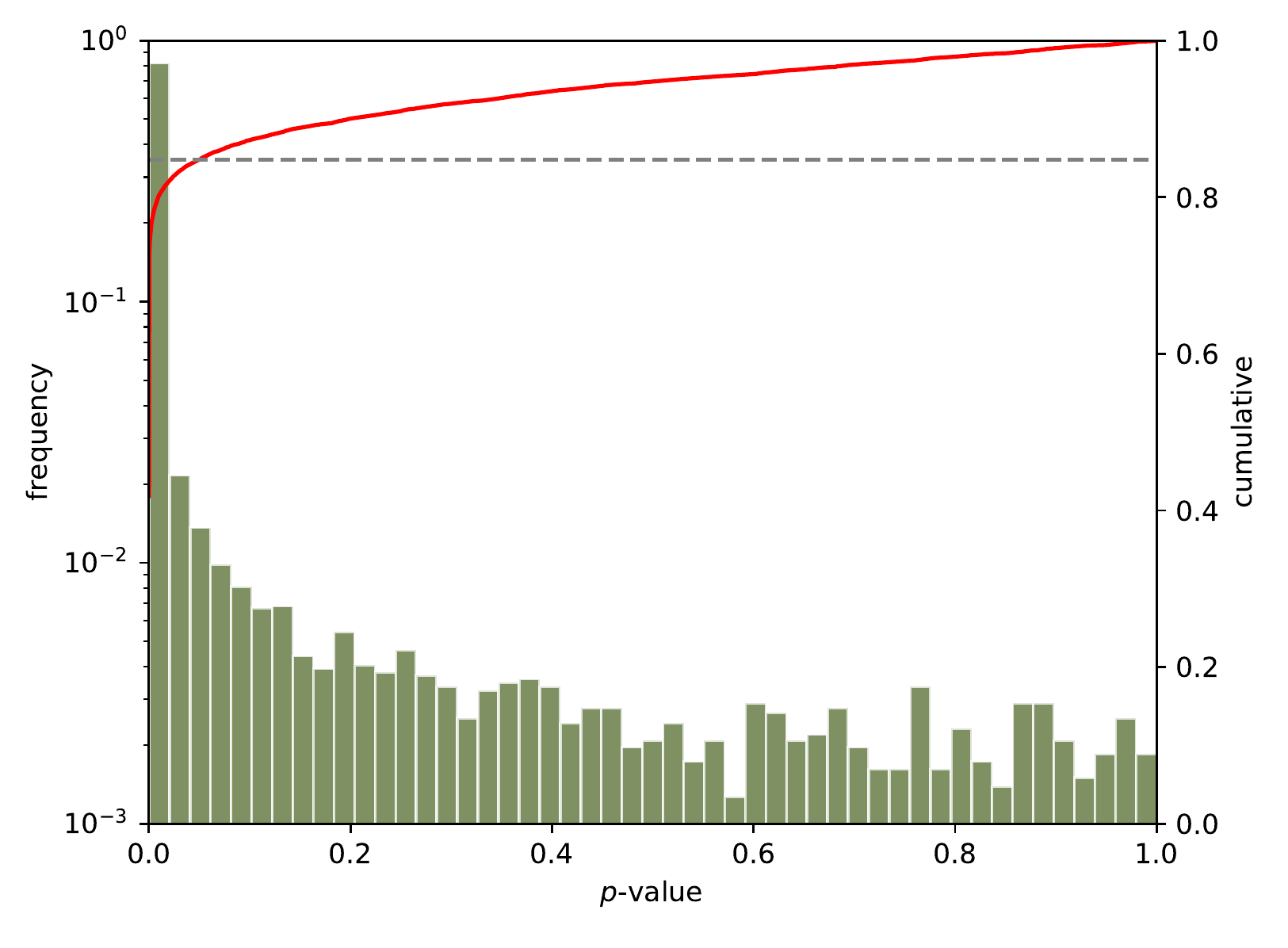}  
 \caption{Histograms of the $p$-values  associated with the collective estimates $\langle g \rangle$ for  the  $8650$ experiments of the FRBP   database.  The null hypothesis is the unbiased estimates assumption, equation (\ref{up}).  The red curve is the cumulative distribution and the  horizontal dashed  line indicates the value of the cumulative distribution at $p = 0.05$. }  
\label{fig:A5}  
\end{figure}

Figure \ref{fig:A4} shows the results  for the case that all forecasts of the FRBP database are clumped together and replaced by  unbiased forecasts.  In this case,  the crowd is superior to any individual in $16.3 \%$ of the experiments (viz., those for which $\xi=0$), whereas it is superior to most individuals in $99.8 \%$ of the experiments (viz., those for which $\xi \leq 1/2$). Therefore,  a crowd of  unbiased forecasters exhibits all the exalted attributes of the wisdom of crowds, but a crowd of human experts  does not.  Hence our qualms about the generality and usefulness of that phenomenon.

Although the stark difference between figure \ref{fig:6} and figure \ref{fig:A4} offers evidence that the  forecasts in the FRBP database are biased, it is  worthwhile  to seek a more quantitative confirmation that those forecasts are in fact biased.  This can be easily achieved if one considers the null hypothesis as the unbiased estimates assumption for which the collective prediction $\langle g \rangle_u$ is drawn from the Gaussian  distribution  (\ref{up}). Hence, the probability that the unbiased estimates assumption  produces  a collective estimate  at least as extreme as the collective estimate $\langle g \rangle $  actually observed in a particular experiment is simply
\begin{equation}\label{pvalue}
p =  1 - \mbox{erf} \left ( \frac{ \mid<g>-G \mid}{\sqrt{2\delta/N}} \right ),
 \end{equation}
 which is the two-tailed $p$-value. Since there is a $p$-value for each experiment, in figure \ref{fig:A5} we present a histogram of the $p$-values associated with the observations $\langle g \rangle $  for  the $8650$ experiments of the FRBP forecast database. The results show that $85\%$ of the experiments have $p$-value less than $0.05$, which confers  statistical significance to the claim that  a significant fraction of the  forecasts in the FRBP database are biased.

\section*{Appendix B}\label{AppB}

\renewcommand{\thefigure}{B\arabic{figure}}
\renewcommand{\theequation}{B\arabic{equation}}
\setcounter{figure}{0}
\setcounter{equation}{0}

\begin{figure}
\centering  
 \includegraphics[width=.47\textwidth]{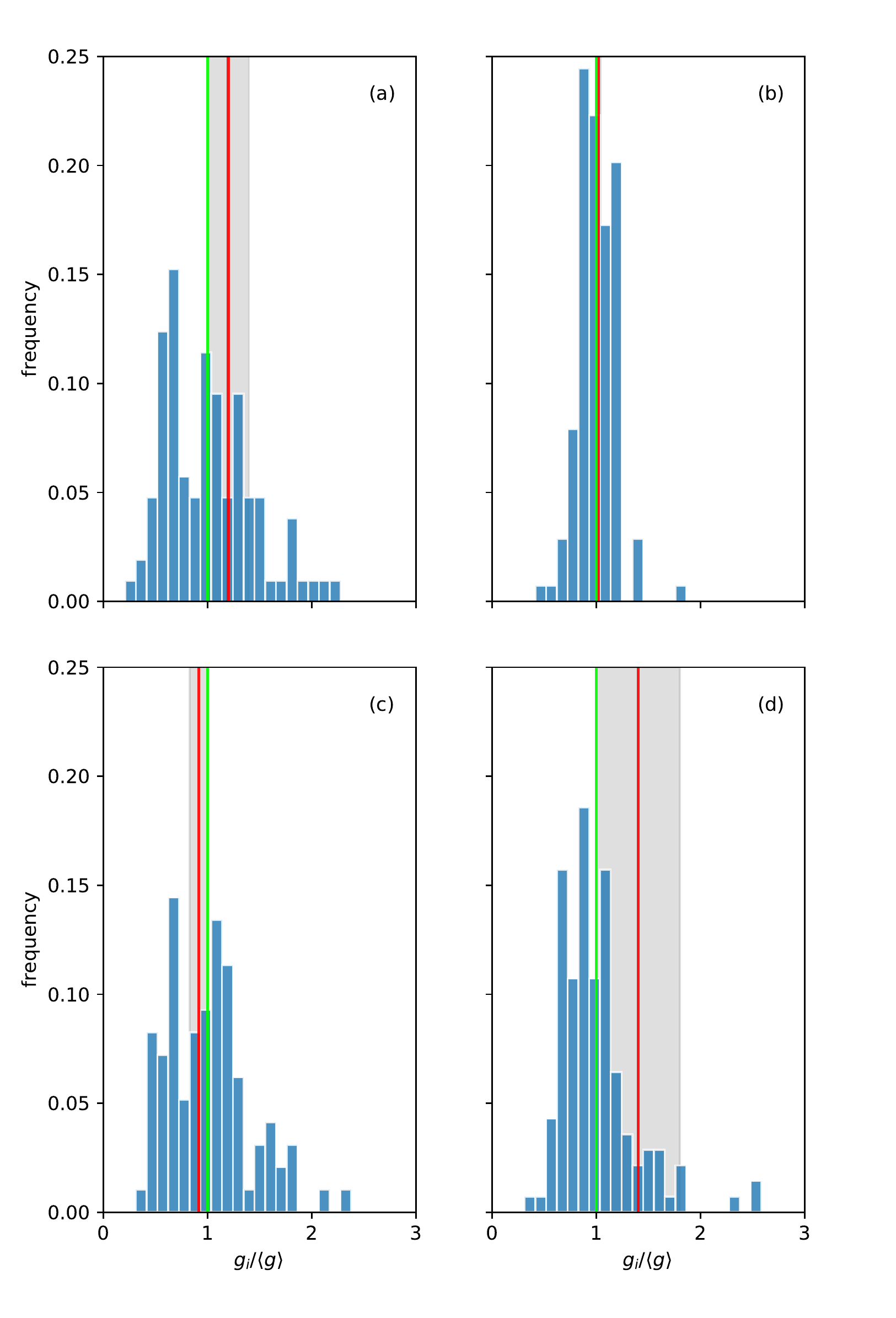}  
 \caption{Histograms of the relative estimates $g_i/ \langle g \rangle$ for  (a) the number of candies in a jar, (b) the length of a paper strip, (c) the weight of a bag of beans and (d) number of pages of a book. The number of participants $N$ is different for each  experiment. The vertical red lines indicate the ratio between the true value and the crowd estimate, i.e., $G/\langle g \rangle$, whereas the vertical green lines at 
 $g_i/ \langle g \rangle = 1$ indicate the scaled  crowd estimate. The gray regions highlight the individual estimates that are better than the crowd's.}  
\label{fig:B1}  
\end{figure}

In this appendix, we analyze briefly   four  wisdom-of-crowds experiments that, in contrast to the experiments discussed in the main text,    counted with the participation of laypeople only.  The experiments are  the estimates of the number of candies in a jar, the length of a paper strip, the number of pages of a book and the weight of a bag of beans. The first three experiments are discussed at length in Ref. \cite{Davi_20}.  The data of the four experiments are publicly available  in \cite{github}. The forecasters are students who have never trained for those kind of guessing tasks and so,  in that sense, they can be  considered as laypeople.  Their skills, however, may vary considerably across the different tasks.  Figure \ref{fig:B1} shows the histograms  of the relative estimates $g_i/ \langle g \rangle$ with $i=1, \ldots, N$  for those experiments.  The number of participants $N$ is different for each  experiment, as specified below.

We begin our analysis with the popular wisdom of crowds experiment  in which  people guess  the number of candies in a jar. In this particular experiment, $N=105$ students  were asked to guess the number of candies in a transparent jar that held $G=636$ candies.
 The resulting distribution of individual estimates  is shown in the  upper left panel of figure \ref{fig:B1}.
  The collective estimate  $\langle g \rangle = 531$ is better than 70\% of the individual estimates. The gray region in the histogram indicates the individual estimates that are closer to the true value $G$  than the collective estimate $\langle g \rangle$. The diversity or variance of the estimates is $\delta = 48736$. The scaled quantities introduced in the main text are $\delta^{1/2}/\langle g \rangle = 0.42$ and $\gamma^{1/2}/G = 0.16$.
  Using the formulation presented in \ref{AppA}, we find that the probability $p$ [see equation (\ref{pvalue})] that a group of $N=105$ unbiased estimators  produces a collective estimate  at least as extreme as  $\langle g \rangle $  is less than $10^{-6}$.  Hence, the participants of this experiment  are definitely  not unbiased estimators.

 In the second experiment we discuss here,   $N=139$ students were asked  to size up a paper strip of length  $G = 22.4~ \mbox{cm}  $.  In contrast to the candies-in-a-jar experiment, we do not expect too far off estimates of the strip length since its true value $G$   is very close to the span unit ($ 1 ~\mbox{span} =22.86~ \mbox{cm}$), which is a natural length standard the students were probably aware of. The  distribution of individual estimates  is shown in the  upper right panel of figure \ref{fig:B1}.
The collective estimate is $\langle g \rangle =  22.0~ \mbox{cm} $ and the variance of the estimates is $\delta= 12.42 ~ \mbox{cm}^2$.
The scaled diversity is $\delta^{1/2}/\langle g \rangle = 0.16$ and the scaled collective error is  $\gamma^{1/2}/G = 0.018$.
We note that the collective  estimate  is better than 85\% of the individual estimates and corresponds to a percentage error of only $1.8\%$.  
In addition,  the probability $p$ that a group of $N=139$ unbiased estimators  produces a collective estimate  at least as extreme as  $\langle g \rangle $  is $0.14$, so we cannot reject the null hypothesis  that the forecasts are unbiased. This result supports our point that unbiased forecasters are in fact very skillful estimators.

 In the third experiment,   $N=97$ students were asked  to estimate the weight of a transparent plastic bag  containing  $G = 1.75~ \mbox{kg}$ of beans.  The  distribution of individual estimates  is shown in the  lower left panel of figure \ref{fig:B1}.
The collective estimate is $\langle g \rangle =  1.91~ \mbox{kg} $ and the variance of the estimates is $\delta= 0.58 ~ \mbox{kg}^2$. Hence, 
the scaled diversity is $\delta^{1/2}/\langle g \rangle = 0.40$ and the scaled collective error is  $\gamma^{1/2}/G = 0.091$.
The collective  estimate  is better than 84\% of the individual estimates. 
The probability $p$ that a group of $N=97$ unbiased estimators  produces a collective estimate  at least as extreme as  $\langle g \rangle $  is $0.03$.

 Finally, in the fourth experiment,   $N=140$ students were asked  to estimate
 the number of pages of a book of $G = 784$ pages.   The  distribution of individual estimates  is shown in the  lower right panel of figure \ref{fig:B1}.
The collective estimate is $\langle g \rangle =  560 $  and the variance of the estimates is $\delta= 40332 $.
The scaled diversity is $\delta^{1/2}/\langle g \rangle = 0.36$ and the scaled collective error is  $\gamma^{1/2}/G = 0.29$, which corresponds to a percentage error of  $29\%$.
The collective  estimate  is better than 62\% of the individual estimates.   
The probability $p$ that a group of $N=140$ unbiased estimators  produces a collective estimate  at least as extreme as  $\langle g \rangle $  is less than $10^{-6}$. Surprisingly, the students turned out  to be very bad estimators of the number of pages in a book, perhaps because physical books are no longer part of their  lives.

Although in these four experiments the crowd estimate turned out to be  better than  the estimates of the majority of the participants, in all them there was a fair chance that a randomly chosen participant beat the crowd. 
For example, in the paper-strip experiment, for which the collective estimate was highly accurate, the crowd was outperformed by $15\%$ of the individual estimates whereas in the  pages-of-a-book experiment it was outperformed by $38\%$ of the participants. We note that the probability of choosing four experiments at random in the FRBP  database such that the crowd beats the majority of the forecasters is $(0.7)^4 \approx 0.24$, so the results of this appendix do not oppose those of the main text.

 Picking the candies-in-a-jar and the paper-strip experiments as representative of tasks where, by design, the participants have low  and high skills, respectively,  we reach the  unsurprising conclusion  that the less skilled the participants are in a given task, the greater the percentage collective error and the variance of their estimates. This  accords with our findings in the main text where the difficulty of the task is associated with the  range of the forecasts. In addition, the unbiased estimates assumption (i.e., the null hypothesis) cannot be rejected in the case the task is such that  the participants are likely to deliver accurate estimates, as in paper-strip experiment.  However, it can safely be rejected in the case that the participants deliver far off estimates as in the  candies-in-a-jar experiment.  Again, these findings agree with the view of unbiased forecasters as ultimate experts.

\end{document}